\documentclass{webofc}
\usepackage[varg]{txfonts}   
\usepackage{xcolor}
\newcommand{\pp}           {pp\xspace}

\newcommand{\PbPb}         {\mbox{Pb--Pb}\xspace}

\newcommand{\pPb}          {\mbox{p--Pb}\xspace}
\newcommand{\AuAu}         {\mbox{Au--Au}\xspace}

\newcommand{\pt}           {\ensuremath{p_{\rm T}}\xspace}

\newcommand{\nineH}        {$\sqrt{s}~=~0.9$~Te\kern-.1emV\xspace}
\newcommand{\seven}        {$\sqrt{s}~=~7$~Te\kern-.1emV\xspace}
\newcommand{\twoH}         {$\sqrt{s}~=~0.2$~Te\kern-.1emV\xspace}
\newcommand{\twosevensix}  {$\sqrt{s}~=~2.76$~Te\kern-.1emV\xspace}
\newcommand{\five}         {$\sqrt{s}~=~5.02$~Te\kern-.1emV\xspace}
\newcommand{\twosevensixnn}{$\sqrt{s_{\mathrm{NN}}}~=~2.76$~Te\kern-.1emV\xspace}
\newcommand{\fivenn}       {$\sqrt{s_{\mathrm{NN}}}~=~5.02$~Te\kern-.1emV\xspace}

\newcommand{\MeVc}         {Me\kern-.1emV/$c$\xspace}
\newcommand{\TeV}          {Te\kern-.1emV\xspace}
\newcommand{\GeV}          {Ge\kern-.1emV\xspace}
\newcommand{\MeV}          {Me\kern-.1emV\xspace}
\newcommand{\GeVmass}      {Ge\kern-.2emV/$c^2$\xspace}
\newcommand{\MeVmass}      {Me\kern-.2emV/$c^2$\xspace}

\newcommand{\ee}           {\ensuremath{e^{+}e^{-}}}

\newcommand {\dphi}        {\ensuremath{\Delta\varphi}\xspace}

\newcommand{\pT}{\ensuremath{p_\mathrm{T}}}

\newcommand{\pTjet}{\ensuremath{p_\mathrm{T}^\mathrm{jet}}}

\newcommand{\gevc}{\ensuremath{\mathrm{GeV/}c}}

\newcommand{\RAA}{\ensuremath{R_\mathrm{AA}}}

\begin{document}
\title{Experimental Status of Jets in Heavy-Ion Collisions}
%
%

\author{\firstname{Jaime} \lastname{Norman}\inst{1}\fnsep\thanks{\email{jknorman@liverpool.ac.uk}} 
}

\institute{University of Liverpool, Oliver Lodge Laboratory, Oxford St, Liverpool L69 7ZE
          }

\abstract{%
  Jet quenching has been one of the most important indicators that ultra-relativistic heavy-ion collisions produce a deconfined state of quarks and gluons, known as the Quark-Gluon Plasma. While the quenching of jets traditionally refers to the energy loss of high-momentum partons, the study of jet quenching has grown into a multi-pronged field where the measurement of jets and their modification in heavy-ion collisions is used as an important tool to study many aspects of QCD deconfinement. This contribution reviews the current experimental status of jets at the LHC and RHIC, and reports recent experimental highlights.
}
\maketitle
\section{Introduction}
\label{intro}

The collision of ultra-relativistic heavy-ion collisions at the LHC and RHIC generates temperatures hot enough to create a deconfined state of quarks and gluons, the Quark-Gluon Plasma (QGP). Measurements up to now have enabled a detailed study of the QGP, which has been determined to exist as a low-viscocity, collectively-expanding, strongly-interacting fluid. The nature of the fundamental degrees of freedom within the QGP, and how a strongly-interacting fluid emerges from the asymptotically free gauge theory of QCD, is however an open question.

One of the most powerful probes of the QGP at a range of length scales are QCD jets - high-energy partons (quarks or gluons) which are observed as a high-energy `spray' of hadrons.
Jets are produced in hard-scattering processes at the start of particle collisions, and their production in vacuum (in $\ee/ep$ collisions) is well understood in QCD. This makes them a useful, `calibrated' probe with which the QGP can be studied, where the in-medium evolution and parton shower of the jet is modified at all stages of the QGP lifetime.
The study of jets in heavy-ion collisions aims to address some of the most pressing questions in the study of the different phases of nuclear matter: what are the emergent ‘bulk’ properties of QCD matter at high temperatures? By what physical mechanisms does a partonic ‘probe’ interact with this matter? Can the individual degrees of freedom of deconfined QCD matter be resolved with jets, and what is their nature?

Theoretical description of the propogation of jets through the QGP requires a consistent description of the jet production, the parton shower and its thermalisation in the medium, parton-medium interactions, plus relativistic hydrodynamic evolution of the medium and its response to the propagating jet. An overview of jet quenching theory and recent developments can be found in~\cite{LA-theory}.

\section{Experimental considerations}
\label{sec-1}

The most significant challenge in measuring jets in heavy-ion collisions is the huge `underlying event' created in these collisions, i.e. particles coming from sources uncorrelated to the hard scattering in which the jet is produced. This leads to a given fraction of the jet transverse momentum ($\pT$) originating from these uncorrelated sources which must be corrected for. In addition, at small jet energies it is expected that a non-negligible fraction of jets contain constituents which do not originate at all from any hard scattering, but instead are made up of completely uncorrelated sources (also referred to as `fake jets' or `combinatorial jets'). Techniques have been developed to correct for both cases.

The underlying event can be subtracted from the jet using an event-by-event procedure. ALICE generally estimates the underlying event of a jet as the median underlying event density of the full event~\cite{ALICE:2012nbx}. Other approaches (see e.g.~\cite{CMS:2021vui}) take into account the event-plane dependence of the underlying event due to flow effects. 
It has also recently been shown that the underlying event density resolution can be improved by calculating it on a jet-by-jet basis, using a regression Neural Network trained on simulated jets embedded into a heavy-ion background~\cite{Haake:2018hqn}; A preliminary measurement from this method is shown in section~\ref{sec:inclusive}.

Combinatorial jets can be suppressed in different ways. The most simple is to impose a jet $\pT$\, or leading hadron $\pT$ cut to suppress the combinatorial background. One drawback here is that this does restrict measurements to higher $\pT$ jets, or bias the fragmentation pattern of the jet, respectively. To push to lower $\pT$, statistical techniques have been developed to subtract the uncorrelated background. One is a `mixed event' technique, where tracks are randomly selected from many events to generate a mixed event background which is subtracted from the measured distribution(s) (see e.g.~\cite{STAR:2017hhs,CMS:2021otx}). Another technique has been developed where the difference is taken between two `triggered' distributions containing the same uncorrelated background component (in particular, a trigger(hadron/$\gamma/\pi^0$)-normalised recoil jet distributions in a lower trigger $\pT$ interval is subtracted from the same distribution in a higher trigger $\pT$ interval, see e.g.~\cite{ALICE:2015mdb}, also the hadron-jet measurement in section~\ref{sec:semiinclusive}).

In order to compare to theory, one has to factor into account that reconstructed jet observables are smeared by detector inefficiencies and resolution effects (in heavy-ion and \pp collisions). In heavy-ion collisions, jets are additionally smeared due to underlying event fluctuations. Nowadays, most measurements correct for these effects by unfolding the detector level quantities to particle level, using statistical unfolding techniques.

\section{Results}
\label{sec:results}

Presented in the following are experimental highlights from the LHC and RHIC. The measurements are grouped into three broad catagories: inclusive jet measurements, semi-inclusive jet measurements and jet substructure. Searches for jet quenching effects in small systems are then briefly discussed.

\subsection{Modification of inclusive jets}
\label{sec:inclusive}

\begin{figure}[]
	\centering
	\includegraphics[width = 0.48\textwidth,clip]{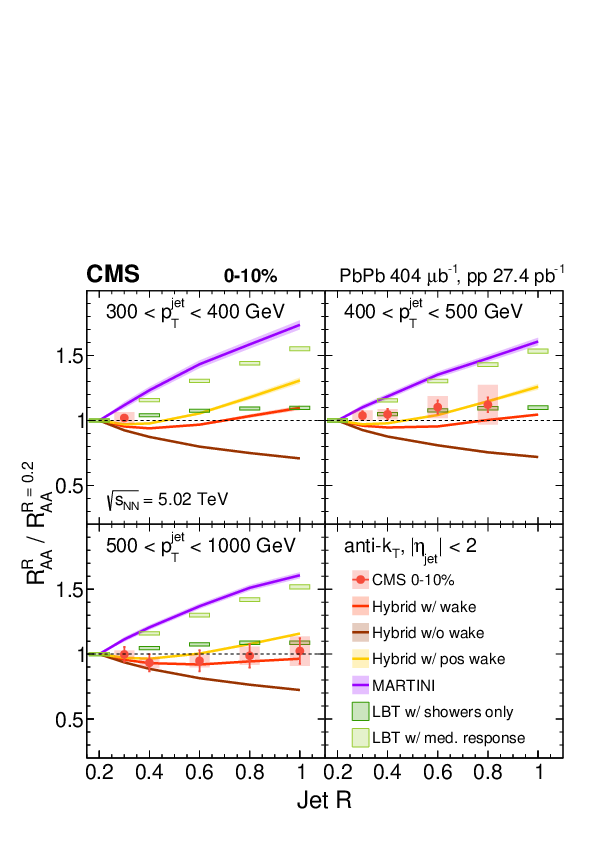}
	\includegraphics[width = 0.48\textwidth,clip]{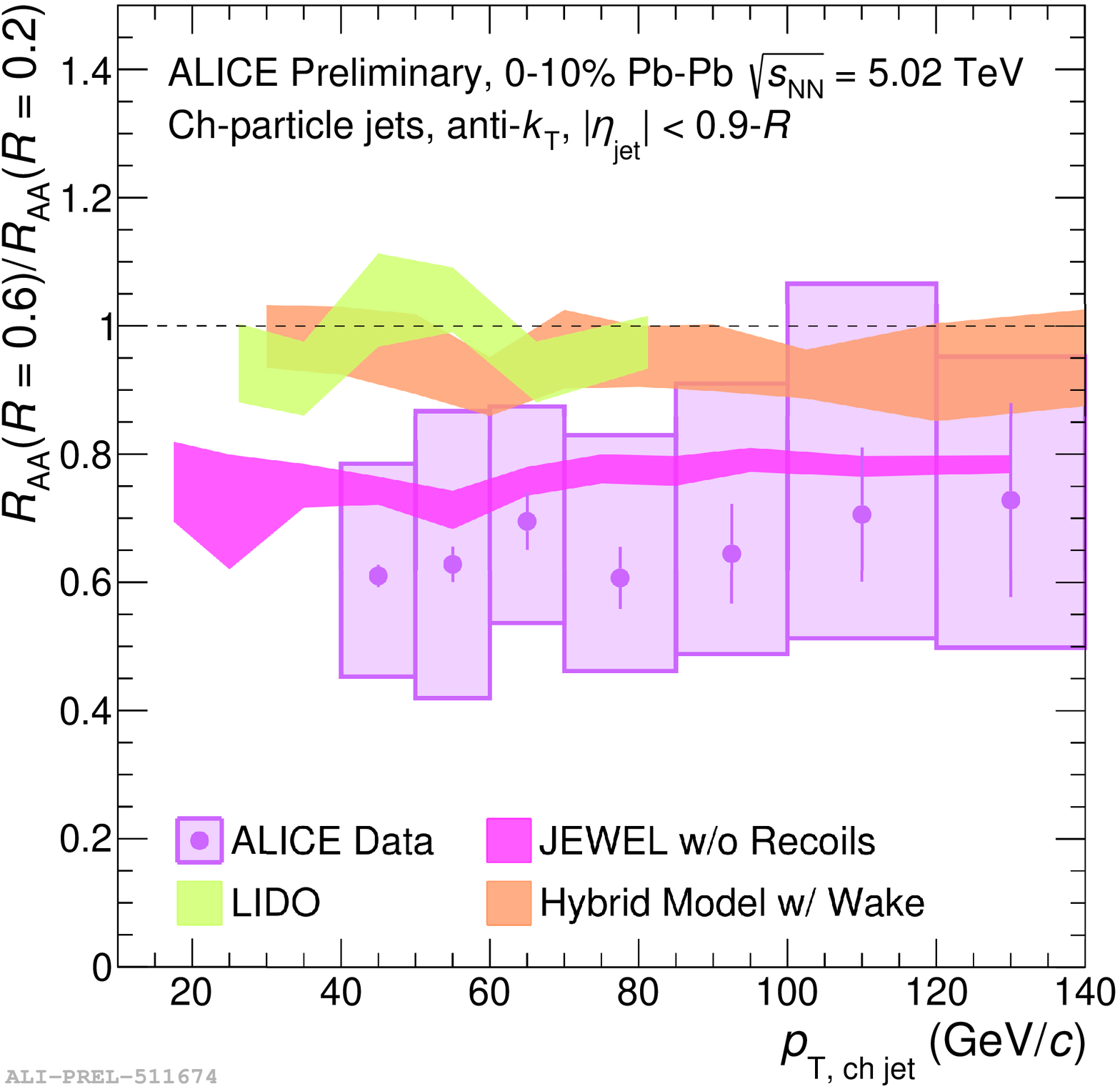}
	\caption{Left: Ratios of the nuclear modification factor \RAA~for different jet radii $R$, in \pTjet~regions 300-400~\gevc, 400-500~\gevc~and 500-1000~\gevc, measured by CMS in central 0-10\% \PbPb collisions~\cite{CMS:2021vui}. Right: The ratio of the nuclear modification factor $\RAA (R = 0.6) / \RAA (R = 0.2)$, measured by ALICE in central 0-10\% \PbPb collisions.}
	\label{fig:RAA}       
\end{figure}

The nuclear modification of jets has been measured by CMS~\cite{CMS:2021vui}, ATLAS~\cite{2019108} and ALICE~\cite{ALICE:2019qyj} at the LHC, and STAR~\cite{STAR:2020xiv} at RHIC. Measurements have shown jets to be suppressed in AA collisions with respect to \pp collisions, indicating the jet loses significant energy outside the jet cone. A suppression persists up to a $\pTjet$ of $\sim1$~\TeV/$c$ in central \PbPb collisions~\cite{2019108,CMS:2021vui}. How this energy is redistributed can be studied further by measuring the $R$ (jet radius parameter) dependence of jet production. Figure~\ref{fig:RAA} (left) shows the ratios of the jet nuclear modification factor $\RAA$~\footnote{The nuclear modification factor $\RAA$ is defined as the ratio of jet yields in AA collisions with respect to \pp collision, $\RAA = \frac{dN^{AA} / d\pT}{\langle N_{coll}\rangle dN^{pp}/d\pT}$, where the \pp yields are scaled by the average number of binary nucleon-nucleon collisions in the collision $\langle N_{coll}\rangle$. An $\RAA<1$ indicates a suppression in AA collisions with respect to pp collisions.} for different jet $R$ measured by CMS as a function of the jet $R$ in the numerator of the ratio~\cite{CMS:2021vui}. The increased statistics in this dataset allowed for measurements at higher $\pT$ and also larger jet $R$ (up to $R=1.0$) than previous measurements. This measurement indicates that the amount of jet suppression has minimal dependence on $R$. ALICE has measured large-$R$ jet production (up to $R=0.6$) at lower $\pT$ than ATLAS and CMS - figure~\ref{fig:RAA} (right) shows the ratio of the nuclear modification factors $\RAA(R=0.6) / \RAA(R=0.2)$, where a hint of a stronger suppression of larger $R$ jets is seen. 
It is noted however that these measurements display tension with an ATLAS measurement~\cite{ATLAS:2012tjt} which shows less suppression for larger $R$ jets, and further measurements are needed to resolve this tension.
Jets tagged for their beauty flavour content have been measured by ATLAS~\cite{ATLAS:2022fgb}, which suggest that the $\RAA$ for $b$-jets is larger than that for inclusive jets, indicating that $b$-jets lose less energy than inclusive jets. A recent measurement by ALICE of charm-tagged jets also shows that there is a hint that charm-jets lose less energy than inclusive jets.

\begin{figure}[]
	\centering
	\includegraphics[width = 0.48\textwidth,clip]{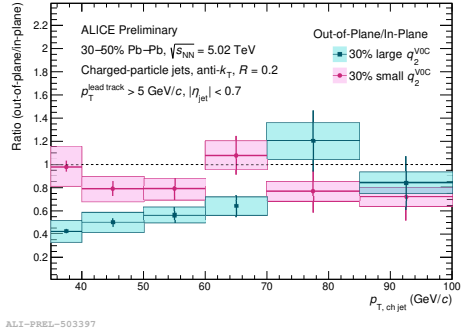}
	\caption{Ratio of $R = 0.2$ jet yields measured in-plane and out-of-plane, for high and low $q_2$ events, measured in semi-central $30-50\%$ Pb--Pb collisions by ALICE.}
	\label{fig:ESE}       
\end{figure}

Measurements can be performed aiming to study the path-length and event-shape dependence of jet production. The centrality dependence of jet azimuthal anisotropy $v_n$ has been measured by ATLAS for the 2nd, 3rd and 4th order cumulants~\cite{ATLAS:2021ktw}, where the $v_{2,3,4}$ values follow a similar trend with centrality for measurements of $v_n$ that are driven by hydrodynamics, indicating that event geometry plays a significant role in jet quenching. CMS measured a positive dijet $v_2$ which follows a similar trend to the inclusive jet $v_2$ with centrality, and a $v_3$ and $v_4$ consistent with 0~\cite{CMS:2022nsv}. ALICE performed a new measurement where events are classified by centrality and their anisotropy. This is done by calculating the 'second-order harmonic reduced flow vector' $q_2 =  | (\sum^M_{i=1} \cos( 2\phi_i ) , \sum^M_{i=1} sin( 2\phi_i ) )| / \sqrt{M}$, where $M$ is the event multiplicity, the sum is over all tracks in the event and $\phi_i$ is the azimuthal angle of track $i$. In this case events with large $q_2$ are less anisotropic, and events with small $q_2$ are more anisotropic.
Figure~\ref{fig:ESE} shows the ratio of jet yields measured in-plane and out-of-plane, for high and low $q_2$ events in semi-central ($30-50\%$) collisions. While the ratio is consistent with unity for small $q_2$ (isotropic) events, the ratio is less than unity for large $q_2$ (anisotropic events) indicating that jets are more suppressed out-of-plane with respect to in-plane in highly anisotropic events.

\subsection{Modification of semi-inclusive jets}
\label{sec:semiinclusive}

While inclusive jet measurements are the most straightforward, the drawback is that the initial energy and/or flavour of the jet can remain unconstrained.
More stringent constraints on the initial energy of the jet and flavour of the jet can be obtained using semi-inclusive coincidence measurements - the measurement of jets recoiling from another object determined to originate from the same hard scattering. These measurements also allow to define an axis to study the deflection of jets due to in-medium scattering. These measurements can include:

\begin{itemize}
	\item photon-jet or electroweak boson (W/Z)-jet coincidence: Since photons and electroweak bosons do not interact strongly, they do not lose energy when traversing the QGP, and can thus be used to tag/constrain the momentum transfer of the hard scattering.
	\item hadron - jet coincidence: Here a high-$\pt$ hadron is used as a proxy for a jet, and the recoiling jets are measured. 
\end{itemize}

\begin{figure}[]
	\centering
	\includegraphics[width = 0.55\textwidth,clip]{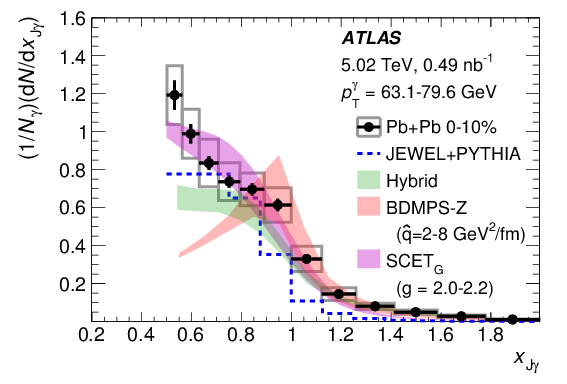}
	\includegraphics[width = 0.44\textwidth,clip]{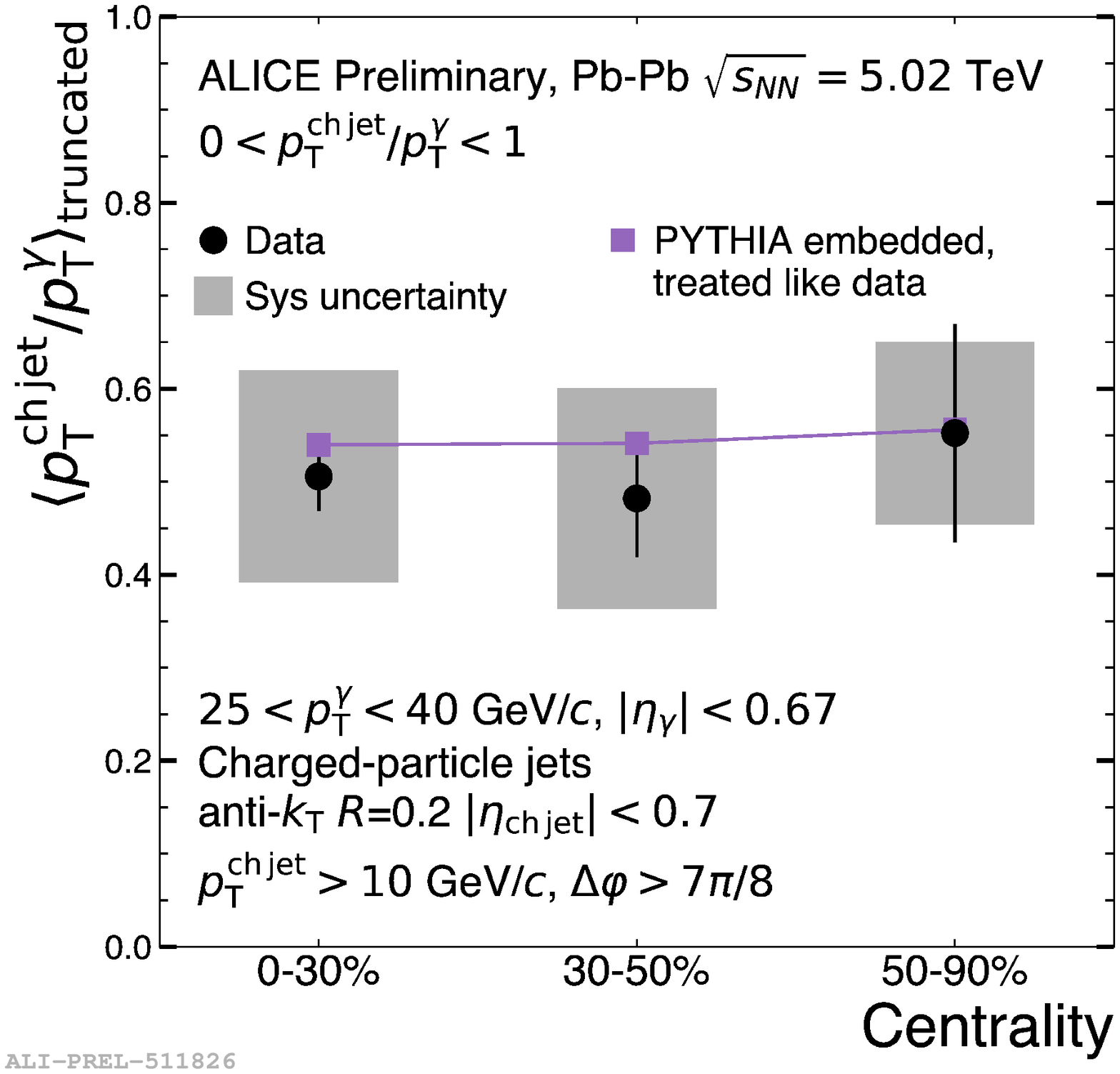}
	\includegraphics[width = 0.48\textwidth,clip]{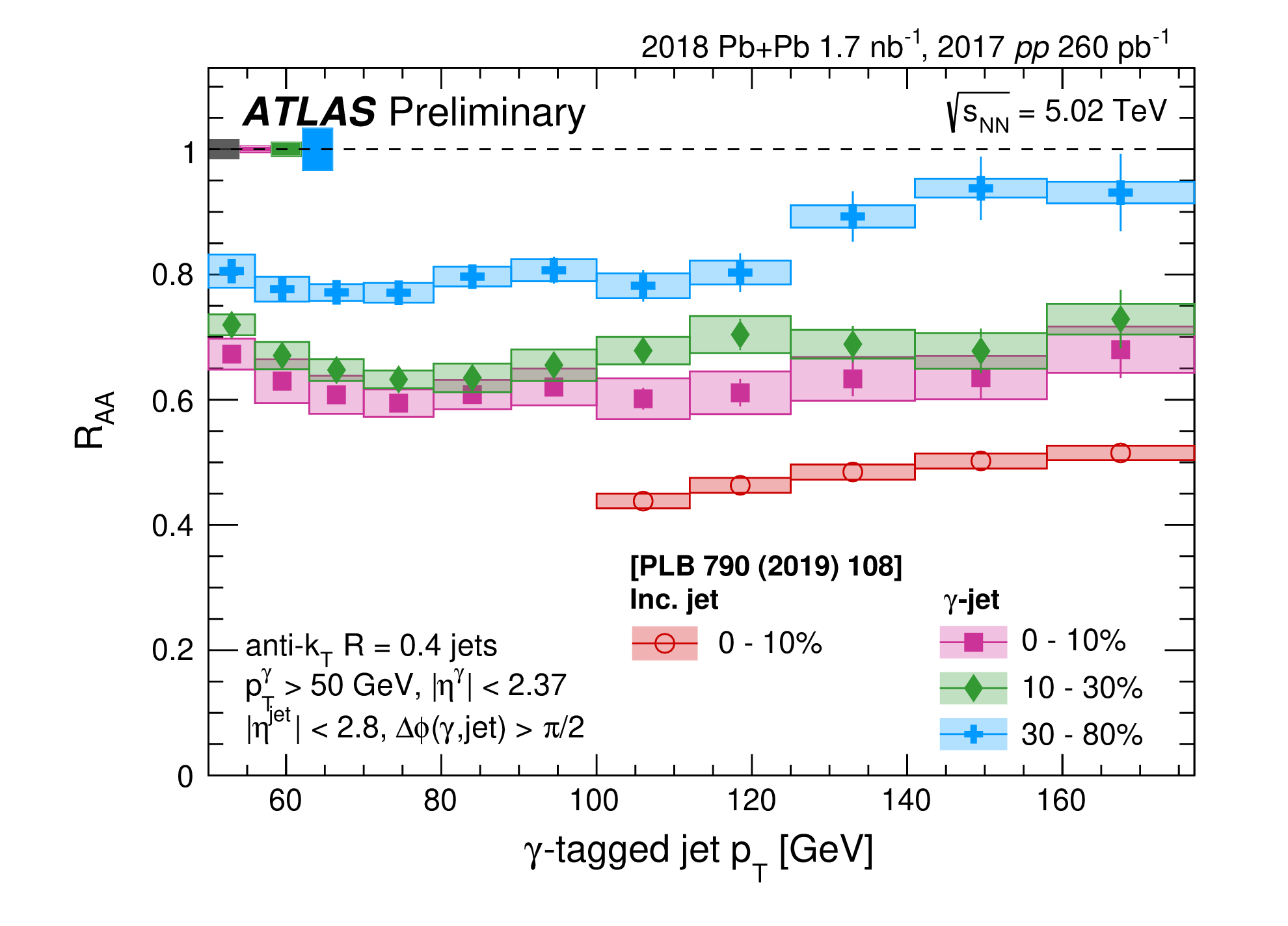}
	\caption{Top left: the momentum imbalance $x_{j\gamma} = \pT^\gamma / \pTjet$ for jets in central 0-10\% \PbPb collisions measured by ATLAS~\cite{ATLAS:2018dgb}. Top right: the truncated mean of the momentum imbalance $x_{j\gamma} = \pT^\gamma / \pTjet$ as measured by ALICE as a function of centrality. Bottom: The $\gamma$-tagged jet $\RAA$ as measured by ATLAS, compared with the inclusive jet $\RAA$. }
	\label{fig:EW-jet}       
\end{figure}

Figure~\ref{fig:EW-jet} (top left) shows the momentum imbalance $x_{j\gamma} = \pT^\gamma / \pTjet$ for $\gamma$-tagged jets measured by ATLAS~\cite{ATLAS:2018dgb}.
CMS measured the momentum imbalance $\pT^\gamma / \pTjet$, azimuthal angle difference $\Delta\phi_{\gamma,jet}$ and $\pTjet$ distribution~\cite{CMS:2017ehl}, where a significant shift in the momentum imbalance in \PbPb with respect to \pp is seen, also indicating significant jet energy loss in \PbPb collisions. ALICE recently measured $x_{j\gamma}$ for lower-\pT\ jets, shown in figure $\ref{fig:EW-jet}$ (top right)(not unfolded due to statistical limitations). In this measurement no significant modification is seen in central collisions with respect to peripheral collisions.
The production of jets in coincidence with photons or electroweak bosons are also significantly more likely to be initiated by a quark (through e.g. Compton scattering $gq \rightarrow q\gamma$) than inclusive jets, which are dominated by gluon-initiated jets, and thus can be used to constrain quark/gluon energy loss. ATLAS recently measured the $\RAA$ of $\gamma$-tagged jets~\cite{ATLAS-CONF-2022-019}, which is shown in figure~\ref{fig:EW-jet} (bottom). $\gamma$-tagged jets are measured to be less suppressed than inclusive jets, indicating quarks lose less energy than gluons.

\begin{figure}[]
	\centering
	\includegraphics[width = 0.7\textwidth,clip]{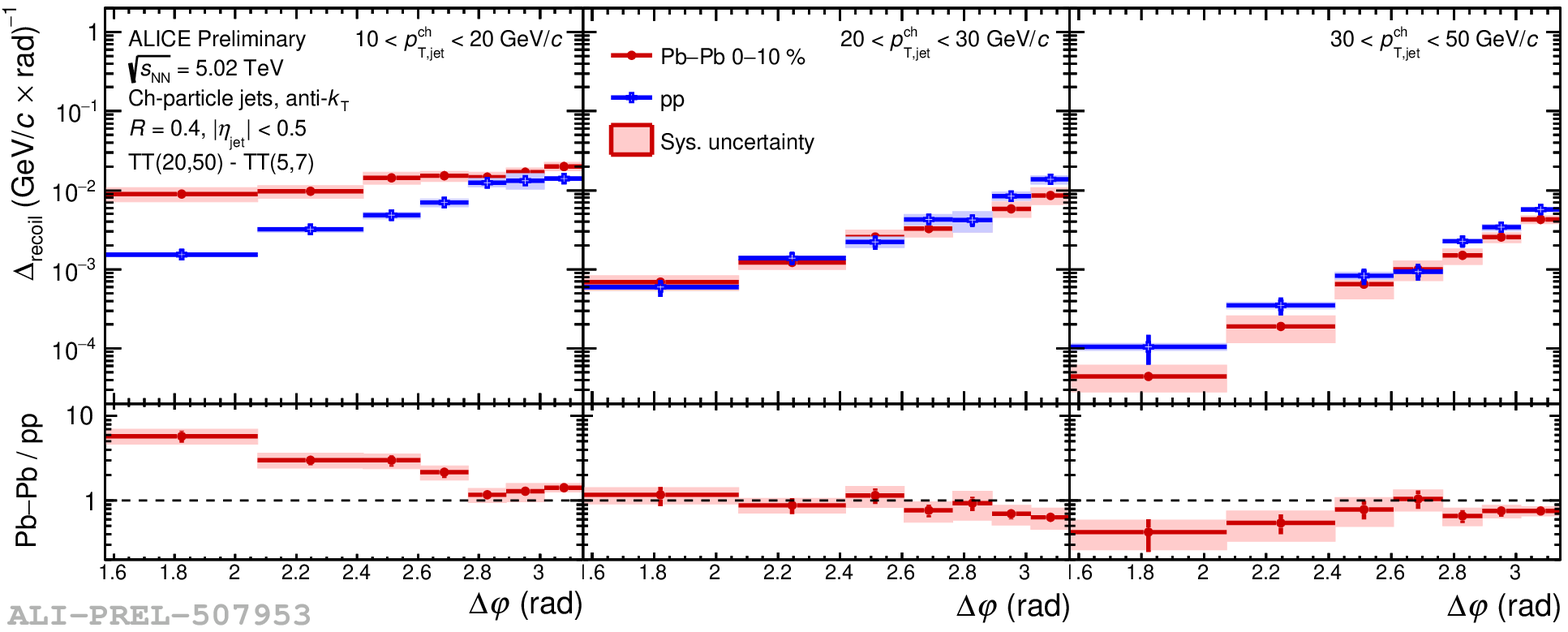}
	\includegraphics[width = 0.29\textwidth,clip]{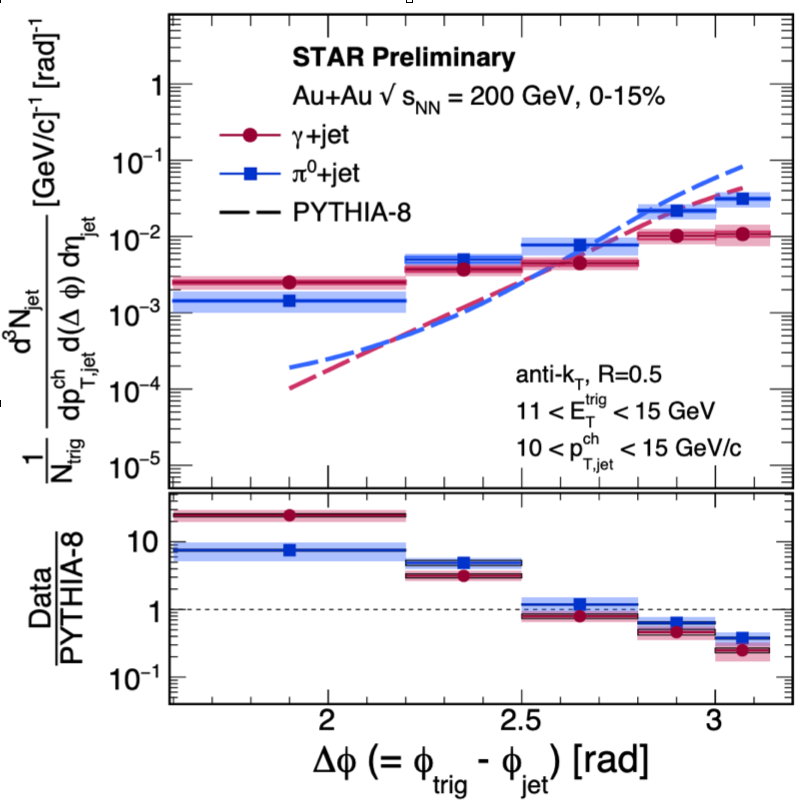}
	\caption{Left: The background-subtracted hadron+jet trigger normalised recoil jet distributions in central 0-10\% \PbPb collisions and \pp collisions, as a function of the azimuthal angle between the trigger and the jet $\dphi$, for different $\pTjet$ intervals, measured by ALICE. Right: The $\gamma$-jet and $\pi^0$-jet trigger-normalised recoil jet distribution as a function of $\dphi$ in central 0-15\% $\AuAu$ collisions, measured by STAR. }
	\label{fig:h-jet}       
\end{figure}

ALICE recently measured jets recoiling from a trigger hadron in \PbPb and \pp collisions, building on previous measurements by ALICE~\cite{ALICE:2015mdb} and STAR~\cite{STAR:2017hhs}. This measurement utilised statistical techniques~\cite{ALICE:2015mdb} to remove the combinatorial jet background, allowing to measure jets down to very low ($\sim 10~\gevc$) \pTjet. Figure~\ref{fig:h-jet} (left) shows the distribution of the azimuthal angle between the trigger hadron and jet for $R=0.4$ jets in selected $\pTjet$ intervals. It is shown that for $10 < \pTjet < 20~\gevc$ jets the $\dphi$ distribution is significantly broadened in \PbPb collisions with respect to \pp collisions. This broadening is accompanied with an overall yield enhancement in the back-to-back region ($|\Delta\varphi - \pi| < 0.6$). The same observation of azimuthal broadening was made by STAR in $\gamma$-jet and $\pi^0$-jet correlations, which is shown in figure~\ref{fig:h-jet} (right).
It is noted that CMS measured the azimuthal angular distribution of $\gamma$-jet correlations~\cite{CMS:2017ehl} for larger jet $\pT$ and found consistency in \pp and \PbPb collisions, further suggesting that the broadening effect occurs just for low-$\pT$ jets.

\subsection{Modification of jet substructure}
\label{sec:substructure}

The time evolution of jets can be visualised as an initial parton-parton scattering, followed by hard parton splittings, followed by successive, softer splittings, finally followed by hadronisation. The study of the substructure of the jet can give important information on how the core of the jet, and how the splittings, are modified by the QGP. Recent developments in jet substructure have developed techniques where a reconstructed jet is reclustered and the clustering procedure is rewound to isolate the `subjets' within the jet. This technique aims to study the jet splittings by filling the `Lund Plane'~\cite{Andersson:1988gp,Dreyer:2018nbf} which is a representation of the relative angular and transverse momentum of a radiative emission with respect to its emitter (here, jet splittings). Grooming techniques can be used to separate out the hard jet core and hard parton splittings from the softened constituents and medium response. These techniques have been proposed~\cite{Andrews:2018jcm} to study different aspects of jet quenching such as whether the subjet structure is resolved by the medium, how the medium response is redistributed, and also how the space-time evolution of the jet is modified.

\begin{figure}[]
	\centering
	\includegraphics[width = 0.52\textwidth,clip]{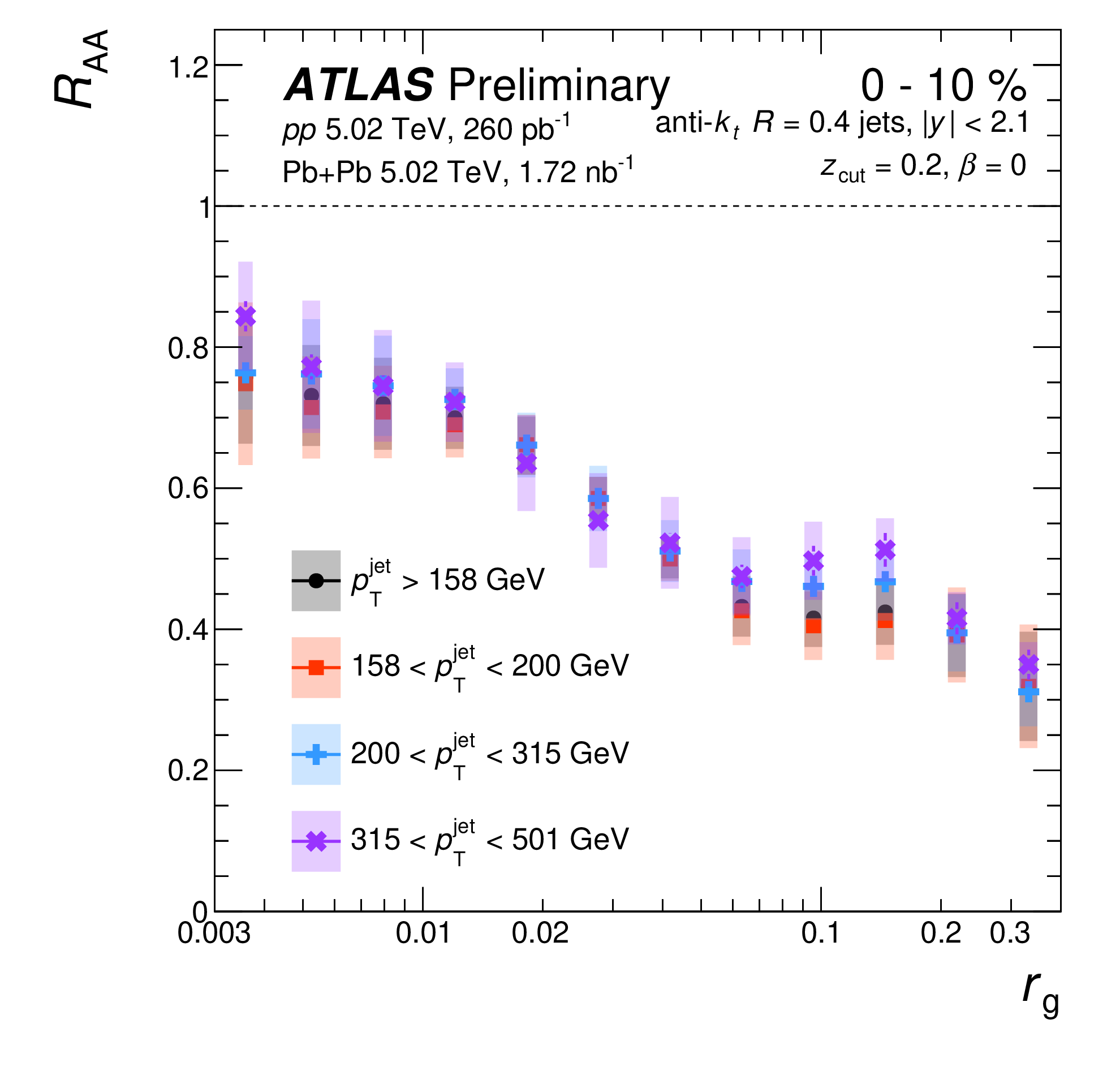}
	\includegraphics[width = 0.47\textwidth,clip]{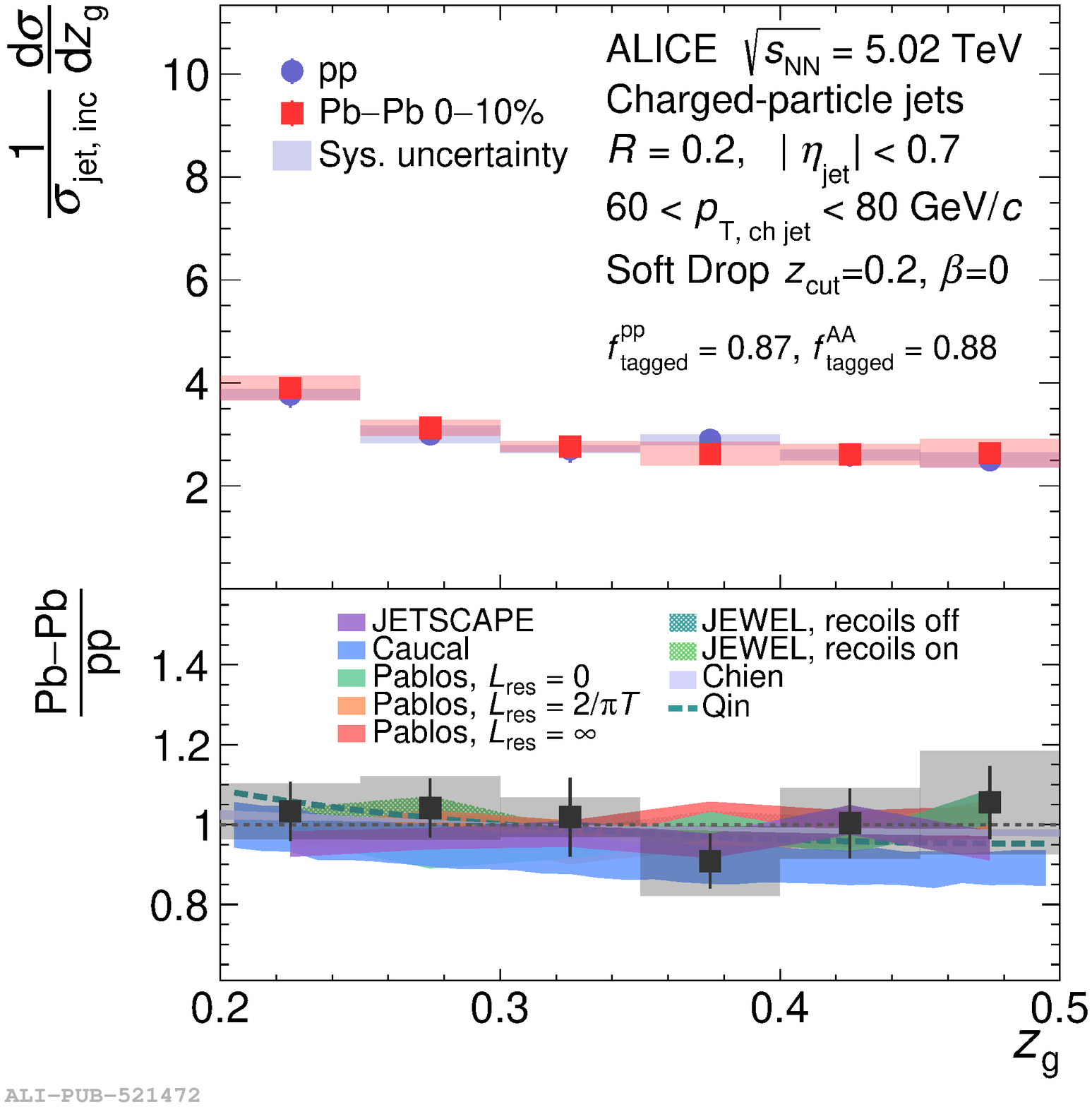}
	\caption{Left: The $\RAA$ as a function of the groomed jet radius $r_g$ in central 0-10\% \PbPb collisions measured by ATLAS~\cite{ATLAS-CONF-2022-026}. Right: The groomed jet momentum splitting fraction $z_g$ for $R=0.2$ jets in central 0-10\% \PbPb collisions and \pp collisions measured by ALICE~\cite{ALargeIonColliderExperiment:2021mqf}.}
	\label{fig:substructure}       
\end{figure}

ATLAS and ALICE have measured the groomed jet radius corresponding to the distance between the leading and sub-leading sub-jets within a jet, $r_g = \sqrt{\Delta\eta_{1,2}^{2} + \Delta\varphi_{1,2}^2 }$~\cite{ATLAS-CONF-2022-026,ALargeIonColliderExperiment:2021mqf}. Shown in figure~\ref{fig:substructure} (left) is the measurement of the $\RAA$ as a function of $r_g$ for groomed jets in four $\pTjet$ intervals by ATLAS. This result shows that jets are narrowed in \PbPb collisions, or wider jets are found to be more suppressed than narrower jets, a result that is also seen by ALICE at lower jet $\pT$. ALICE also measured the groomed jet momentum splitting fraction defined as $z_g = \frac{p_{\mathrm{T,subleading}}}{p_{\mathrm{T,leading}}+p_{\mathrm{T,subleading}}}$, shown in figure~\ref{fig:substructure} (right)~\cite{ALargeIonColliderExperiment:2021mqf}. This result indicates that there is minimal modification to the relative $\pT$ scale of leading and subleading subjets.

\subsection{Search for jet quenching in small systems}
\label{sec:small}

One of the most significant surprises from the LHC physics program is that collective effects traditionally seen as indicative of a deconfined state of quarks and gluons have been measured in smaller collision systems (\pPb and \pp). It is therefore an important question whether the quenching of jets also extends to these collision systems.
Measurements of inclusive jet production~\cite{ALICE:2016faw} and semi-inclusive h+jet production~\cite{ALICE:2017svf} in \pPb collisions as a function of the centrality of the collision, and $b$-jet production in \pPb collisions~\cite{ALICE:2021wct}, do not provide any evidence for jet quenching.
ATLAS recently measured the near-side and away-side yield of charged hadrons correlated with reconstructed jets as a function of the \pPb collision centrality~\cite{ATLAS:2022iyq}. The ratios of these yields in \pp and \pPb collisions $I_{AA}$ is consistent with MC generator AGANTYR which does not include any final state effects producing collectivity or jet quenching, thus also providing no evidence for jet quenching in \pPb collisions.

\section{Summary and Outlook}
\label{sec:summary}

Many new insights into the QGP have been obtained from recent measurements of jets at the LHC and RHIC. Significant modification to jet kinematics and jet substructure have been measured. To date no evidence for jet quenching in small systems (\pp and \pPb) has been seen. The measurement highlights shown in this contribution offer constraining power to theoretical calculations, which use different approaches to calculate jet production and modification in heavy-ion collisions. Recent approaches to estimate transport properties, connecting theory and experiment, have been performed (see e.g. Bayesian parameter estimation of single particle spectra to estimate the jet transport coefficient $\hat{q}$ from JETSCAPE~\cite{JETSCAPE:2021ehl}). Such methods offer a promising way to combine multiple jet measurements and rigorously compare with theory to extract quantitative information about the QGP.

Run 3 at the LHC has begun after major upgrades to the LHC experiments and a heavy-ion run is scheduled for 2023.
At RHIC, sPHENIX is a significant experiment upgrade which will begin collecting data in 2023, complementary to the LHC program. The next few years will thus open up a more precise era in the measurement of jets, allowing to study the QGP with unprecedented accuracy.

\bibliography{references}

\end{document}